\documentclass[runningheads]{cl2emult}

\usepackage{makeidx}  
\usepackage{graphicx} 
\usepackage{subeqnar} 
\usepackage{multicol} 
\usepackage{cropmark} 
\usepackage{eso}      
\usepackage{rotating} 
\makeindex            

\def\Journal#1#2#3#4{{#1} {\bf #2}, #3 (#4)}

\def\APJ{\em Astrophys. Journal}
\def\APJL{\em Astrophys. Journal Lett.}

\def\MNRAS{{\em MNRAS}}
\def\Nat{{\em Nature}}

\def\be{\begin{equation}}
\def\ee{\end{equation}}
\def\bea{\begin{eqnarray}}
\def\eea{\end{eqnarray}}

\newcommand{\bm}[1]{{{\rm\bf #1}}}

\begin{document}
\title*{Noise estimation in CMB time-streams and fast map-making.
 Application to the BOOMERanG98 data.}
\toctitle{Noise estimation and iterative map-making}
%
%
\titlerunning{Noise estimation and iterative map-making}
%
\author{S.Prunet\inst{1}
P.A.R.Ade\inst{2},  
J.J.Bock\inst{3}, 
J.R.Bond\inst{1}, 
J.Borrill\inst{5,6}, 
A.Boscaleri\inst{7}, 
K.Coble\inst{8}, 
B.P.Crill\inst{9},
P. de Bernardis\inst{4},
G. De Gasperis \inst{10}, 
G. De Troia\inst{4}, 
P.C.Farese\inst{8}, 
P.G.Ferreira\inst{11}, 
K.Ganga\inst{9,11}, 
M.Giacometti\inst{4}, 
E.Hivon\inst{9}, 
V.V.Hristov\inst{9}, 
A.Iacoangeli\inst{4}, 
A.H.Jaffe\inst{6}, 
A.E.Lange\inst{9}, 
L.Martinis\inst{13}, 
S.Masi\inst{4}, 
P.Mason\inst{9}, 
P.D.Mauskopf\inst{14}, 
A.Melchiorri\inst{4}, 
L.Miglio\inst{4,15}, 
T.Montroy\inst{8}, 
C.B.Netterfield\inst{15}, 
E.Pascale\inst{7}, 
F.Piacentini\inst{4}, 
D.Pogosyan\inst{1}, 
F.Pongetti\inst{16},
S.Prunet\inst{1}, 
S.Rao\inst{16}, 
G.Romeo\inst{16}, 
J.E.Ruhl\inst{8}, 
F.Scaramuzzi\inst{13}, 
D.Sforna\inst{4}, 
N.Vittorio\inst{10} 
 }

\authorrunning{S. Prunet et al.}
%
%
\institute{
$^1$Dipartimento di Fisica, Universit\'a di Roma La Sapienza, Roma, Italy;
$^2$Dept. of Physics, Queen Mary and Westfield College, London, UK; $^3$Jet 
Propulsion Laboratory, Pasadena, CA, USA; $^4$CITA University of Toronto, Canada;
$^5$NERSC-LBNL, Berkeley, CA, USA; $^6$Center for Particle Astrophysics, 
Univ. of California at Berkeley, USA; $^7$IROE - CNR, Via Panciatichi 64, 50127
Firenze, Italy; $^8$Department of Physics, Univ. of California at Santa Barbara, USA; 
$^9$California Institute of Technology, Pasadena, USA; $^{10}$Dipartimento di 
Fisica, Universit\'a di Roma Tor Vergata, Roma, Italy; $^{11}$Astrophysics, University 
of Oxford, UK; $^{12}$PCC, Coll\`ege de France, Paris, France; $^{13}$ENEA Centro 
Ricerche di Frascati, Italy ; $^{14}$Physics and Astronomy Dept, Cardiff 
University, UK; $^{15}$Departments of Physics and Astronomy, Univ. of Toronto, Canada; 
$^{16}$Istituto Nazionale di Geofisica, Roma, Italy}

\maketitle              

\begin{abstract}
We describe here an iterative method for jointly estimating the 
noise power spectrum from a CMB experiment's time-ordered
data, together with the maximum-likelihood map. 
We test the robustness of this method on simulated Boomerang 
datasets with realistic noise.
\end{abstract}

\section{Introduction}
The map-making problem for CMB anisotropy measurements was first considered
in the context of the COBE-DMR mission \cite{Lin94,Wri94}. 
It was further extended to quick algorithms for differential measurements \cite{Wri96a} 
taking into account $1/f$ noise \cite{Wri96b}.
Given the size of the upcoming datasets ($\geq 10^6$ pixels for {\em MAP}) it is
essential that the map-making algorithm remains quick (typically $O(n_d\log L_N)$
where $n_d$ is the number of time-samples, and $L_N$ the effective noise-filter
length). Another related question, which has been pioneered by \cite{Fer99}, is
how to determine the noise statistical properties from the data itself.
We present a map-making method, based on Wright's fast algorithm, to compute
iteratively a minimum variance map together with an estimate of the detector
noise power spectrum which is needed to compute unbiased power spectrum estimators
of the cosmological signal. 

\section{Iterative map-making - Application to simulations}
\subsection{Method}
\label{sec:method}

We model the data stream in the following way:
\begin{equation}
\bm{d_t} = \bm{P_{tp}\Delta_p} + \bm{n_t}
\end{equation}
where $\bm{\Delta_p}$ is a pixelized version of the {\em observed} sky (i.e. convolved
by the experimental beam), and $\bm{n_t}$ is the detector noise after primary
deconvolution of any filter present in the instrumental
chain (e.g. bolometer time constant, read-out filters). 
We will assume here that the experiment is a total power measurement,
i.e. that the pointing matrix $\bm{P_{tp}}$ contains only one non-zero element
per row. 
We now want to estimate the minimum variance map from this data, i.e. the map
that minimizes $\chi^2 = ({\bm d}-{\bm P\Delta})^\dagger{\bm N}^{-1}({\bm d}-{\bm P\Delta})$.
The solution is given by:
\begin{equation}
\label{eq1}
\tilde\Delta = \left({\bm P}^\dagger{\bm N}^{-1}{\bm P}\right)^{-1}
{\bm P}^\dagger{\bm N}^{-1}{\bm d}
\end{equation}
A few remarks are necessary at this point. First, the matrix to be inverted
is huge ($n_{pix}\times n_{pix}$) so that an iterative linear solver is needed.
Secondly, the noise correlation matrix $N^{-1}_{tt'}$ has to be determined from
the data itself. To make this tractable we assume that, at least over subsets of
the time-stream, the noise is reasonably stationary, so that the multiplication by
$\bm{N}^{-1}$ becomes a convolution operator, in other words that it is diagonal
in Fourier space \footnote{This is actually only approximately true since a convolution
operator is a circulant matrix, i.e. it assumes that the time-stream has periodic boundary
conditions; this is however a rather good approximation for a time-stream much longer
than the effective length of the noise-filter \cite{Teg97}}.
We thus implemented the following algorithm:\\
\begin{center}
\fbox{\hskip 1cm\parbox[c]{8cm}{
{\bf for each stationary noise subset}
\begin{itemize}
\item{$ \bm{n}^{(j)} = \bm{d} - \bm{P\tilde\Delta}^{(j)} \Rightarrow \bm{N}^{(j)-1} = 
\langle\bm{nn}^\dagger\rangle^{-1} $}
\item{$ \bm{\tilde\Delta}^{(j+1)} - \bm{\tilde\Delta}^{(j)} = 
\left(\bm{P^\dagger W^* P}\right)^{-1}\bm{PN}^{(j)-1}\bm{n}^{(j)}$}
\end{itemize}
{\bf endfor}}}
\end{center}
\vskip 0.5cm
To keep the algorithm as fast as possible, we took $\bm{W^*}$ to be diagonal 
and constant, so that $\bm{P^\dagger W^* P}$ is diagonal, with each element
being equal to the number of observations per pixel, up to a multiplicative
constant. The choice of the numerical value of this constant is only important
for the convergence properties of the algorithm. 
In this form, the algorithm is very similar to a Jacobi iterative solver
(in which we would have $\left(diag\{\bm{P^\dagger N}^{-1}\bm{P}\}\right)^{-1} $
instead of $\left(\bm{P^\dagger W^* P}\right)^{-1}$). It has the additional advantage
of being very easy to compute, for very similar convergence properties.
Since we made the assumption that each noise matrix is diagonal in Fourier space,
all time-domain operations are done in Fourier space using FFTs, thus reducing
the number of floating point operations to $O(L_N\log L_N)$ for each subset.

The advantages of this iterative method are obvious: it is fast ($O(n_d\log L_N)$
operations) and cheap in memory ($O(L_N)$ storage). However, since we
try to estimate both the noise power spectrum and the map in a leap-frog manner,
the convergence properties must be studied with numerical simulations.
In particular, the stability of the algorithm is function of the number of
parameters that we use to describe the noise power spectrum. In other words, we have
to assume some regularity in the latter, which translated in practice in fitting
the amplitude of the noise in frequency band powers of varying width.
  
\subsection{Simulations}

To test the properties of the algorithm, we made simulations of Boomerang LDB time-streams
with an sCDM sky,
i.e. with precisely the same scanning strategy and a beam smoothing comparable to
the 150a channel. We chose this channel because it was used to determine 
the angular power-spectrum $C_\ell$ of the CMB anisotropies \cite{Bern00}.
The 10 days of data are approximately divided in two halves corresponding to two
different azimuthal scan speeds (2 degrees, then 1 degree per second) at a mean elevation
of $\sim 45$ degrees. These two scan-speeds
excited differently mechanical/thermal resonances in the instrument payload, 
and therefore were characterized by different noise power-spectra at low frequency;
we simulated the noise time-streams for both halves according to the 
noise power spectra as measured by our method over the 150a data.

\begin{figure}[htb]
\label{fig2}
\hbox{
  \includegraphics[width=0.5\textwidth]{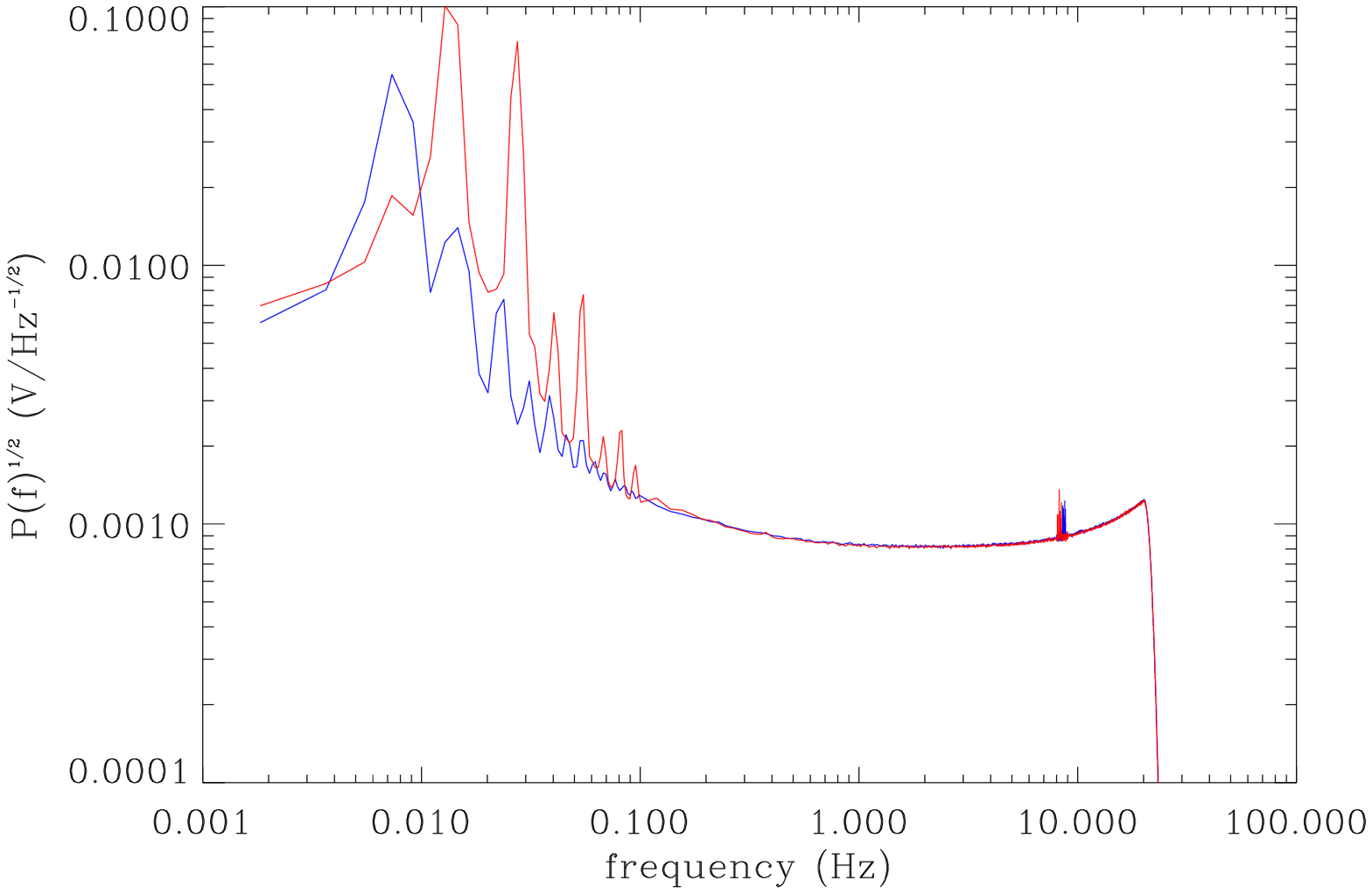}
  \includegraphics[width=0.5\textwidth]{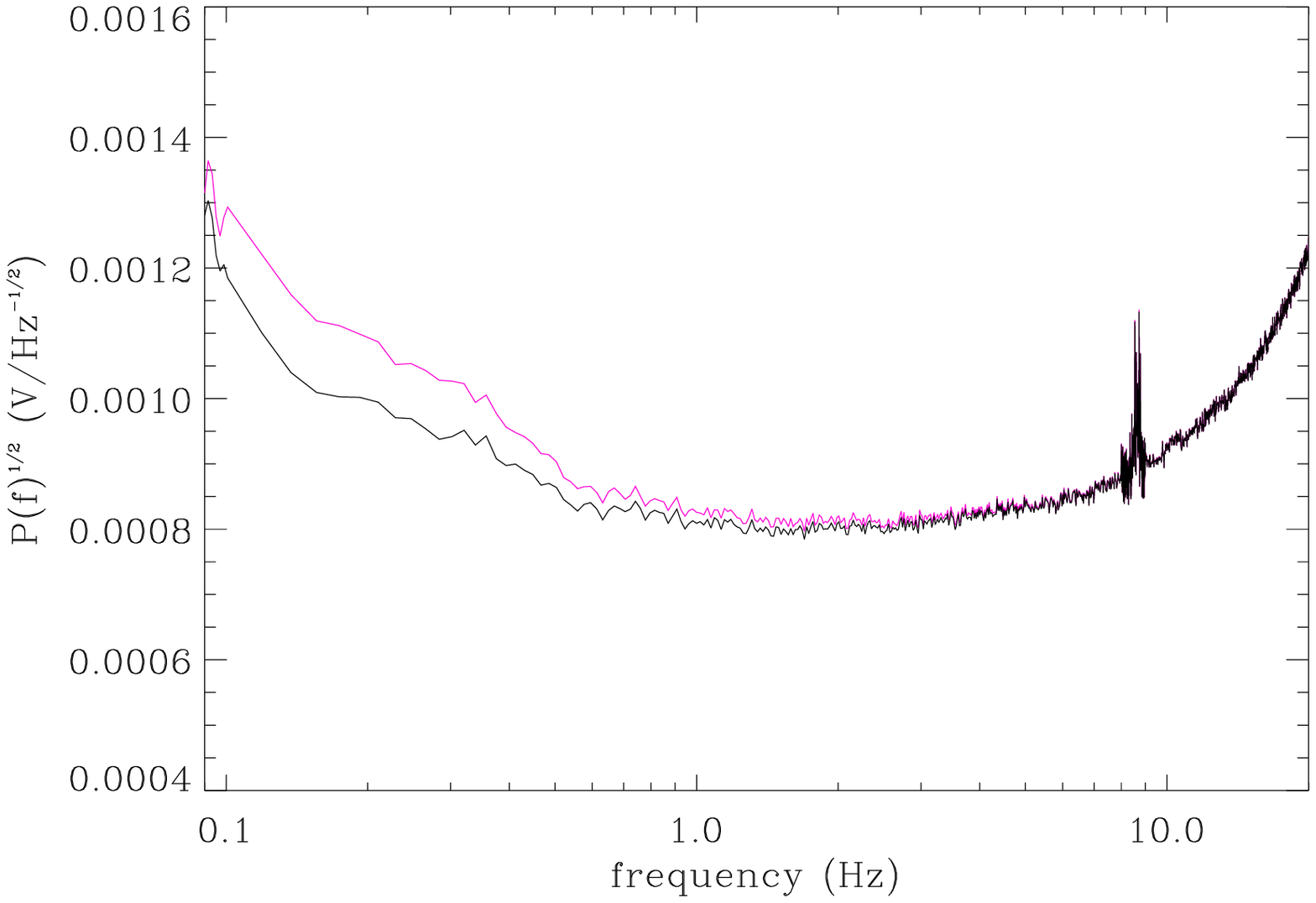}
}
\caption{{\em Left panel:} Time-stream noise spectra of the two halves of the 150a channel data,
estimated with the method described in section\ref{sec:method}. The blue line corresponds
to one degree per second data (1 dps), the red line to 2 dps data. Note the difference
of amplitude in the scan-synchronous lines. {\em Right panel:} Spectrum of the (signal+noise)
1 dps simulated data (magenta) together with the recovered noise power spectrum (black).}
\end{figure}

These power spectra are shown in figure~\ref{fig2} (left panel), for both scan velocities.
One can see in particular that for the two degree per second spectrum (dotted line)
the scan-synchronous harmonics are much stronger than in the one degree per second
spectrum (where the scan-synchronous signal is actually dominated by the dipole
emission). For a more complete description of these effects, as well as a precise
description of the instrument, we refer the reader to Brendan Crill's thesis (\cite{Crill00}).

The right panel of figure~\ref{fig2} shows the
(1 degree per second) spectrum of the simulated (signal+noise) data (magenta)
together with the recovered noise spectrum (black). One can see that most of the signal power
is concentrated in a smooth continuum contribution between $\sim 0.1\,{\rm Hz}$ and 
$\sim 3-4\,{\rm Hz}$
(which corresponds approximately to the beam size). This plot shows that a naive estimation
of the noise from the data time-stream would typically result in overestimating the noise by 
$\sim 10\,\%$  at the degree scale.

In principle, the presence of strong scan-synchronous lines in the noise power spectra 
at low frequency, as well as the overall $1/{\rm f}$ property of the noise could be taken
into account by the method which would down-weight those frequencies accordingly, but in practice,
because of the phase-coherent, non-gaussian nature of those scan-synchronous patterns we had 
to high-pass filter the data at $0.06\times{\rm scan\,speed\,(Hz)}$.
This changes the effective pointing matrix $\bm{P}$ to $\bm{FP}$ 
where $F$ is the high-pass filter; however,
looking at Eq.~\ref{eq1}, this is equivalent to leaving $\bm{P}$ unchanged,
and replacing $\bm{N}^{-1}$ by $\bm{F^\dagger N}^{-1}\bm{F}$, which is straightforward
in Fourier space\footnote{It should be noted here that any {\em invertible} filter applied
to the data does lead to the same maximum likelihood solution, as expected. In the case of
the high-pass filter considered here, it is not invertible, and therefore affects the noise properties
of the resulting map by effectively projecting out some modes in the time-stream}.

The Boomerang data set provides a flag with each sample data, indicating the mode of observation
(CMB scans, source scans, SZ regions, etc.) and the quality of the data (either good or bad, where
bad data corresponds to cosmic ray contamination or a scan elevation
change resulting in fluctuations of the cold plate temperature). We thus flagged as bad in 
our simulations all samples not corresponding to good CMB scan data. 
In principle, the proper way to deal with such a data with gaps would be to construct the orthonormal 
noise eigenmodes basis with support constrained to the valid chunks, and do the whole analysis
on this basis instead of Fourier space. However, given the length of the time-stream considered
here ($\sim 5\times10^7$ time samples) it is completely impractical. The solution we adopted
was to model those gaps as being samples spent on the observation of a $0\,{\rm K}$ internal 
calibrator considered as an additional fake pixel. We had thus to make constrained realizations
of the noise time-stream as measured by the algorithm at each iteration step in those gaps,
and assign their pointing matrix element outside the map to this fake pixel.

Here again, the length of the time-stream comes again as a problem, since the constrained 
realizations involve in the best case (when the separation between gaps is bigger than the
noise correlation length $N_{\rm cor}$) inverting a matrix of linear size $N_{\rm cor}$,
and in the worse case a matrix of linear size comparable to the entire time-stream.
Fortunately, fast algorithms have been developed to answer this problem
(known as linear prediction). In particular, we implemented a version of Burg's algorithm 
(see e.g. \cite{Press92}) to solve this problem. 

\begin{sidewaysfigure}
  \label{fig1}
  \hskip -1.5cm
  \vbox{
    \hbox{
      \includegraphics[width=0.55\textwidth,angle=90]{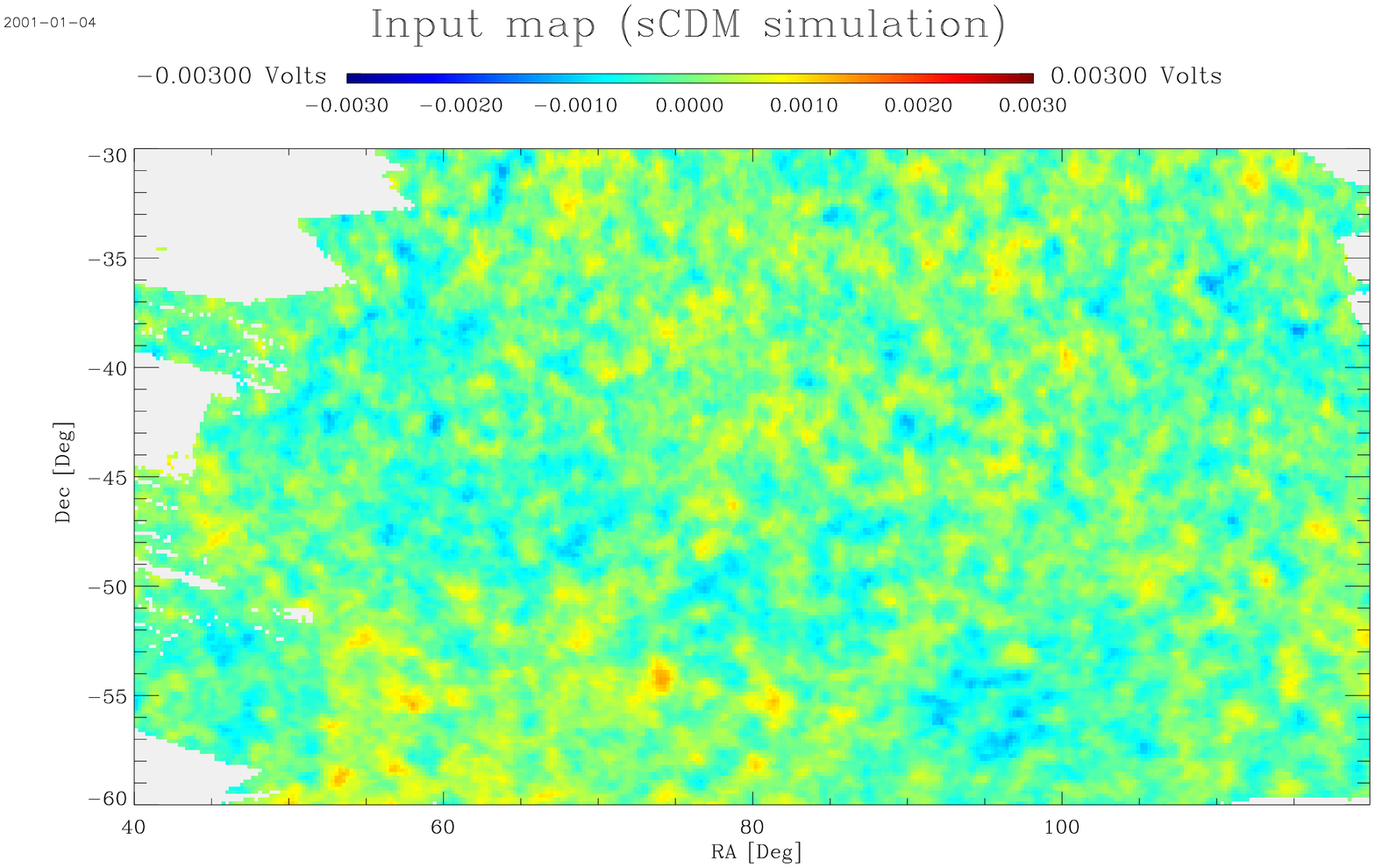}
      \includegraphics[width=0.55\textwidth,angle=90]{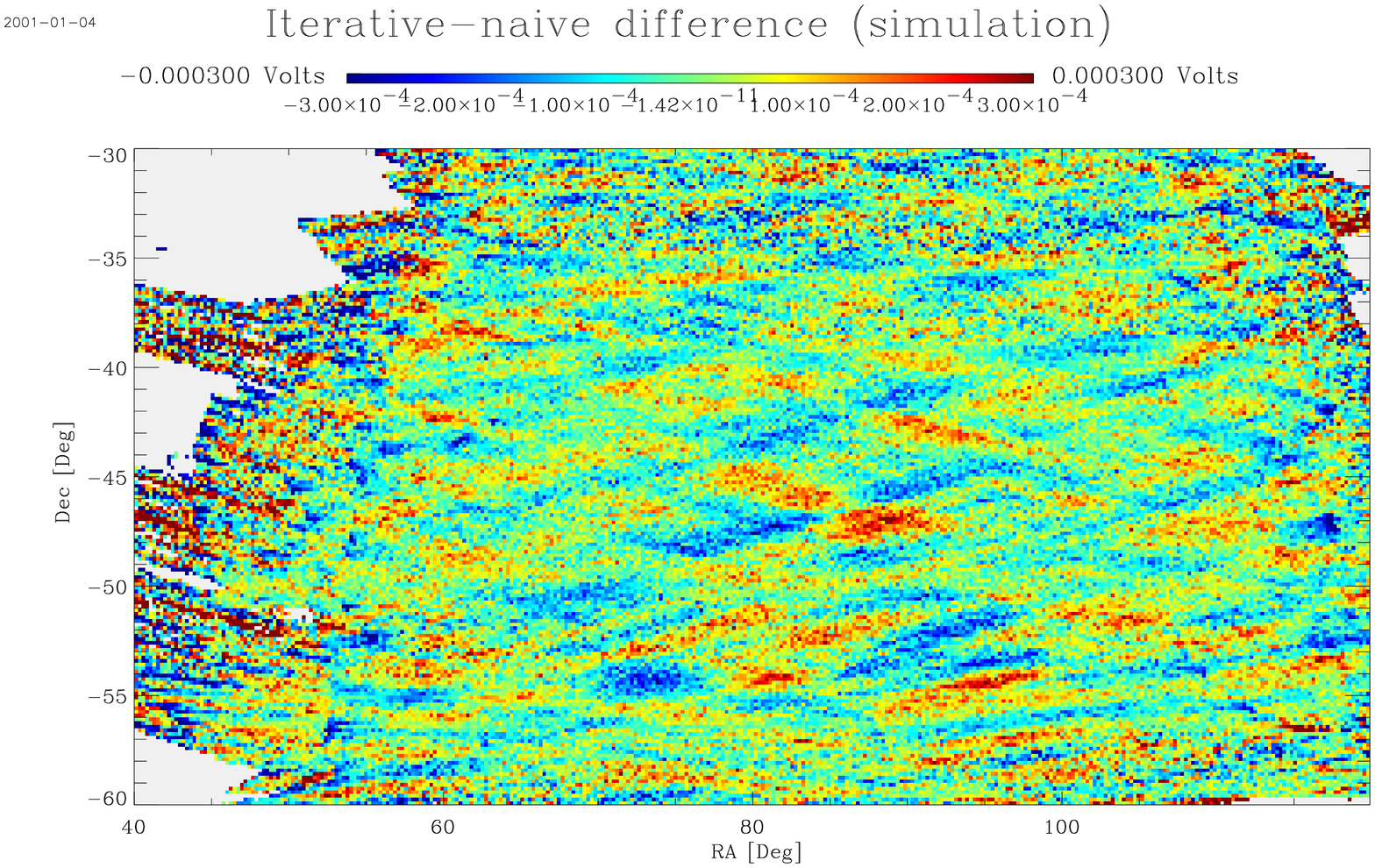}
      }
    \hbox{
      \includegraphics[width=0.55\textwidth,angle=90]{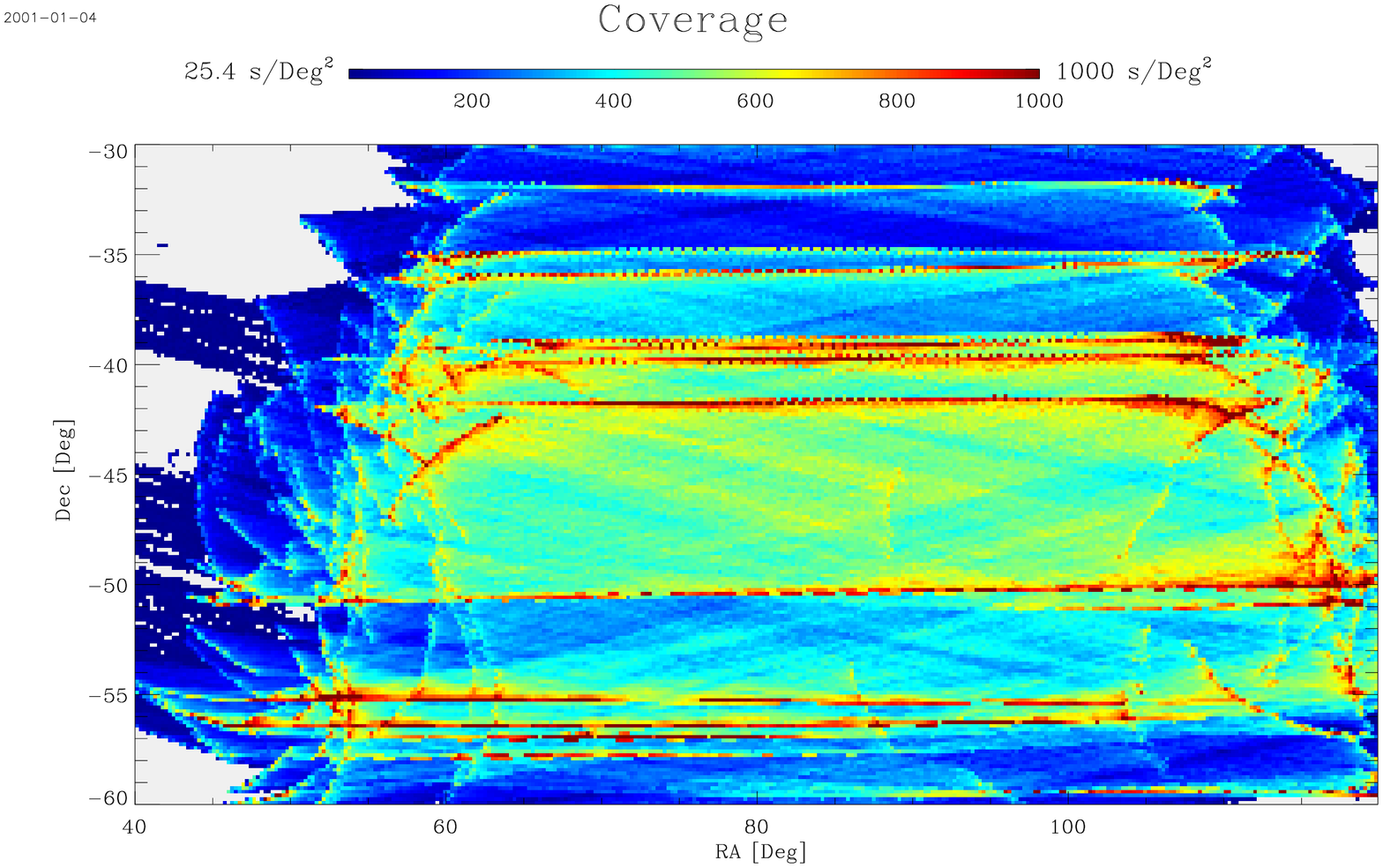}
      \includegraphics[width=0.55\textwidth,angle=90]{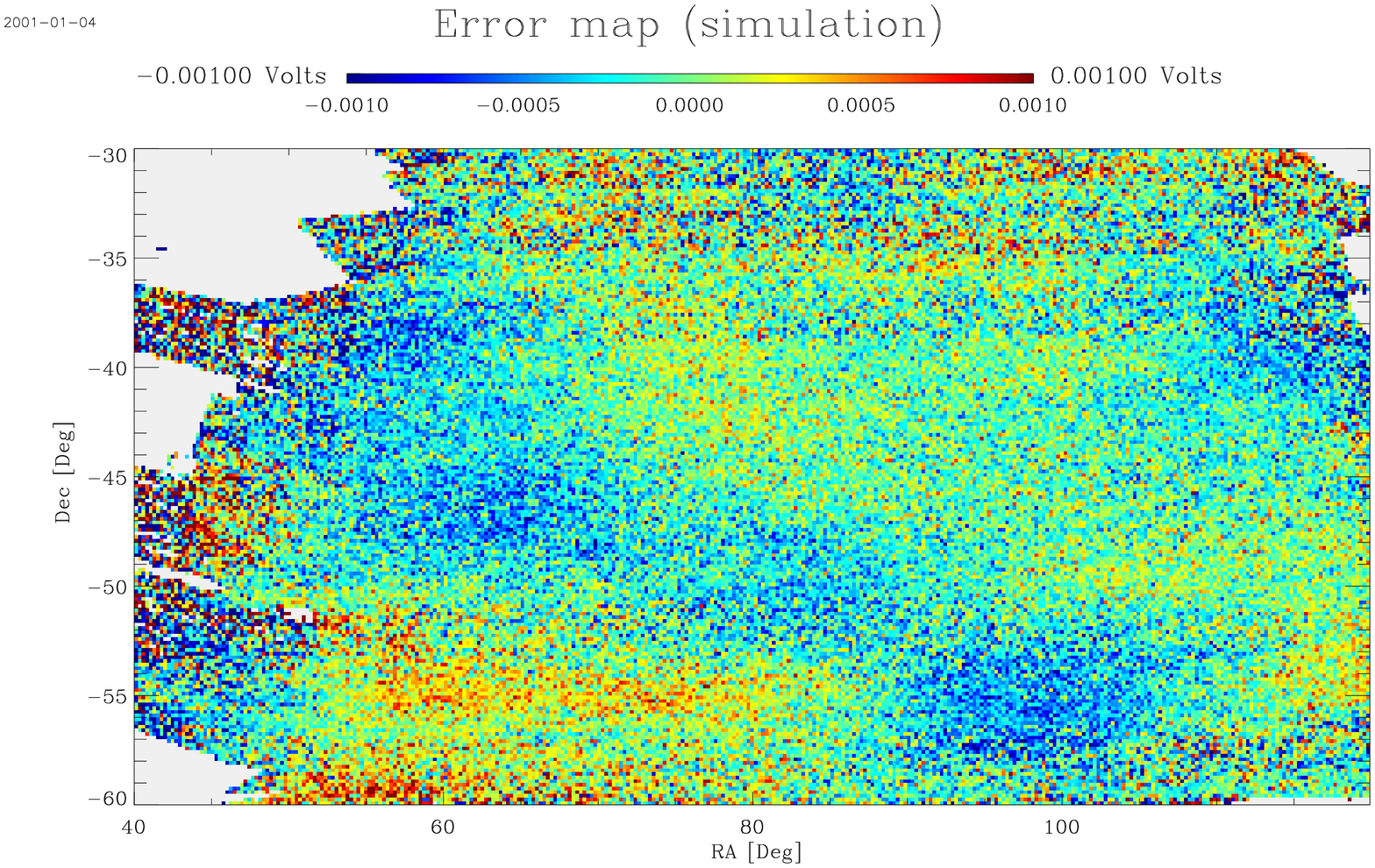} 
     }
    }
  \caption{\small Simulations - Upper left: Input map - Lower left: Coverage 
    - Upper right: Striping map (difference between iterative and coadded map). 
    - Lower right: error map (iterative case). Note the different color scales
    used for the difference maps.}
\end{sidewaysfigure}

Finally, we used HEALpix $7'$ pixels \cite{Gor98}, resulting in a $138620$ pixel map.
The input simulated map (i.e. ``before observation'') is shown in figure~\ref{fig1} 
(upper left panel), together with the corresponding sky coverage (lower left panel).
The different ``broad bands'' in the coverage file correspond to distinct scan periods 
of the flight at different constant sky elevations, 
illustrating the compromise between a uniform sky coverage and a constant elevation 
scanning strategy (to avoid atmospheric gradients).
The lower right panel shows the difference between the input map and the map recovered
from the time-stream by the iterative method. We can see several important things in this
error map, namely:
\begin{itemize}
\item The net effect of the high-pass filtering in the time-stream is to kill the largest
scales in the map. The fact that the pointing matrix has been effectively changed 
by the filtering has been properly taken into account by the algorithm (see discussion
above).
\item There is no apparent residual striping in the error map, although a more precise
way to quantify this statement may be necessary to be able to compare this method with alternative
map-making algorithms.
\item The apparent horizontal features in the error map are related to non-uniform coverage
(compare to coverage map in the lower-left panel), and are {\em not} caused by residual
striping. 
\end{itemize}
Residual striping can be seen in the upper right panel of figure~\ref{fig1},
where the difference of the coadded (``naive'') and iterative maps is shown. We can in particular
see that the features follow the mean direction of the scan (as expected from striping) and
form a non-zero angle to the iso-dec direction.

The method is not free of caveats however. 
As in any iterative linear solver, the convergence of the solution, decomposed
on the (noise) kernel eigenmodes, is a function of the associated eigenvalue.
Thus the noisiest pixel modes (usually at large scales since the time-stream
is high-pass filtered, either by hand or as part of the algorithm if the noise
is higher at low frequency) take (exponentially) more time to converge.
The problem scales as the noise matrix condition number, which is a direct function of
the noise power spectrum dynamical range and of the scanning strategy.
A solution to this problem \cite{Dore00} is a multi-grid version of the present
algorithm, exploiting the hierarchical structure of the HEALPix
pixelization, and rebinning the time-stream accordingly. \cite{Dore00} have shown 
that the convergence rate on large scales is then drastically reduced.
Another caveat is that the distribution of the noise power spectrum estimation is
skewed, leading to a small bias in the noise estimation. This is similar
to the ``cosmic bias'' discussed by \cite{BJK00} in the case of the $C_\ell$'s.
A possible solution would be to approximate the likelihood distribution of the noise
spectrum estimator to correct for this bias. 
Finally, this method does not deal yet with beam asymmetry, but 
fast convolution methods \cite{Wan00} are being developed, and will be ultimately
integrated, allowing the addition of different channels with different beams
to the same map.

\section{Conclusions - Perspectives}
  

We described a fast iterative map-making method
to simultaneously generate the maximum-likelihood map and the noise power spectrum
from a scanning experiment time stream. We tested its convergence properties 
on realistic Boomerang simulations, including the effect of complex scan strategy
and noise power spectra,
resulting in a complicated noise matrix.
We concluded that, except for the spatially largest 
(and ill conditioned) modes, the map and noise power spectra converge
very quickly. This caveat can be cured by means of a ``convergence accelerator'',
where multi-grid methods provide a very promising solution.
In addition, this method, coupled to a fast $C_\ell$ quadratic estimator \cite{Wan00b,Sza00}, 
provide a practical way to compute the angular power spectrum of CMB fluctuations
in the future megapixel experiments.

\clearpage
\flushbottom


\begin{thebibliography}{7}
%
\addcontentsline{toc}{section}{References}

\bibitem{BJK00} J.R. Bond, A.H. Jaffe, L. Knox, \Journal{\APJ}{533}{19}{2000}

\bibitem{Bern00} P. de Bernardis {\em et al}, \Journal{\Nat}{404}{955}{2000}

\bibitem{Crill00} B.P. Crill, Ph.D. thesis, CalTech. (2000)

\bibitem{Dore00} O. Dor\'e, this volume

\bibitem{Fer99} P.G. Ferreira \& A.H. Jaffe, \Journal{\MNRAS}{312}{89}{2000}

\bibitem{Gor98} K.M. G\'orski, E. Hivon, B.D. Wandelt in proceedings
of the MPA/ESO Conference, Garching, 2-7 August 1998, eds A.J. Banday, 
R.K. Sheth and L. Da Costa. See also {\rm http://www.tac.dk/~healpix/}

\bibitem{Lin94} C.H. Lineweaver {\em et al}, \Journal{\APJ}{436}{452}{1994}

\bibitem{Press92} W.H. Press, B.P. Flannery, S.A. Teukolsky, W.T. Vetterling, 1992,
{\em Numerical Recipes}, Cambridge University Press

\bibitem{Sza00} I. Szapudi, S. Prunet, D. Pogosyan, A. Szalay \& J.R. Bond, astro-ph/0010256 

\bibitem{Teg97} M. Tegmark, \Journal{\APJL}{480}{87}{1997}

\bibitem{Wan00} B.D. Wandelt, this volume

\bibitem{Wan00b} B.D. Wandelt, E. Hivon, K.M. Gorski, astro-ph/0008111

\bibitem{Wri94} E.L. Wright, in proceedings of the BCSPIN Puri Winter School

\bibitem{Wri96a} E.L. Wright, G. Hinshaw, C.L. Bennett, \Journal{\APJL}{458}{53}{1996}

\bibitem{Wri96b} E.L. Wright, proceeding of the IAS CMB Workshop, astro-ph/9612006  

\end{thebibliography}
\end{document}